\documentclass[conference]{IEEEtran}
\ifCLASSOPTIONcompsoc
\else
\fi

\ifCLASSINFOpdf
\else
\fi

\usepackage{cite}
\usepackage{url}
\usepackage{minted}
\usepackage{wrapfig}
\usepackage[pdftex]{graphicx}
\usepackage{float}
\usepackage{algorithmic}
\usepackage[tight,footnotesize]{subfigure}
\usepackage{booktabs}
\usepackage{caption}
\usepackage{adjustbox}
\usepackage{makecell}
\usepackage[linesnumbered,ruled]{algorithm2e}
\usepackage{listings}
\usepackage{xcolor}
\usepackage{enumitem}
\usepackage{pifont}
\newcommand{\cmark}{\ding{51}}%
\newcommand{\xmark}{\ding{55}}%
\usepackage[left=1.62cm,right=1.62cm,top=1.9cm]{geometry}

\definecolor{codegreen}{rgb}{0,0.6,0}
\definecolor{codegray}{rgb}{0.5,0.5,0.5}
\definecolor{codepurple}{rgb}{0.58,0,0.82}
\definecolor{backcolour}{rgb}{0.95,0.95,0.92}

\begin{document}


\title{SRE-Llama --- Fine-Tuned Meta's Llama LLM, Federated Learning, Blockchain and NFT Enabled Site Reliability Engineering(SRE) Platform for Communication and Networking Software Services}


 
\author{
\IEEEauthorblockN{
Eranga Bandara\IEEEauthorrefmark{1},
Safdar H. Bouk\IEEEauthorrefmark{1},
Sachin Shetty\IEEEauthorrefmark{1},
Ravi Mukkamala\IEEEauthorrefmark{1},
Abdul Rahman\IEEEauthorrefmark{2},
Peter Foytik\IEEEauthorrefmark{1},\\
Ross Gore\IEEEauthorrefmark{1},
Xueping Liang\IEEEauthorrefmark{3},
Ng Wee Keong\IEEEauthorrefmark{4},
Kasun De Zoysa\IEEEauthorrefmark{5}
}

\IEEEauthorblockA{
\IEEEauthorrefmark{1}
{\rm\{cmedawer, sbouk, sshetty, mukka, pfoytik, rgore \}@odu.edu}\\
Old Dominion University, Norfolk, VA, USA\\
}
\IEEEauthorblockA{\IEEEauthorrefmark{2}
{\rm abdulrahman@deloitte.com}\\
Deloitte \& Touche LLP \\
}
\IEEEauthorblockA{\IEEEauthorrefmark{3}
{\rm xuliang@fiu.edu}\\
Florida International University, USA\\
}
\IEEEauthorblockA{\IEEEauthorrefmark{4}
{\rm AWKNG@ntu.edu.sg }\\
Nanyang Technological University, Singapore\\
}
\IEEEauthorblockA{\IEEEauthorrefmark{5}
{\rm\{kasun\}@ucsc.cmb.ac.lk}\\
University of Colombo School of Computing, Sri Lanka\\
}
}

\maketitle

\begin{abstract}




Software services are crucial for reliable communication and networking, therefore, Site Reliability Engineering (SRE) is important to ensure these systems stay reliable and perform well in cloud-native environments. SRE leverages tools like Prometheus and Grafana to monitor system metrics, defining critical Service Level Indicators (SLIs) and Service Level Objectives (SLOs) for maintaining high service standards. However, a significant challenge arises as many developers often lack in-depth understanding of these tools and the intricacies involved in defining appropriate SLIs and SLOs. To bridge this gap, we propose a novel SRE platform, called ``SRE-Llama'', enhanced by Generative-AI, Federated Learning, Blockchain and Non-Fungible Tokens (NFTs). This platform aims to automate and simplify the process of monitoring, SLI/SLO generation, and alert management, offers ease in accessibility and efficy for developers. The automation processes are governed by smart contracts on the Blockchain, ensuring transparency and security. The system operates by capturing metrics from cloud-native services and storing them in a time-series database, like Prometheus and Mimir. Utilizing this stored data, our platform employs Federated Learning models to identify the most relevant and impactful SLI metrics for different services and SLO objectives values, addressing concerns around data privacy and decentralized data sources. Subsequently, custom-trained Meta's Llama-3 LLM is adopted  to intelligently generate SLIs, SLOs, Error-budgets, and associated alerting mechanisms based on these identified SLI metrics. The Llama-3-8B LLM has been quantized and fine-tuned using Quantized Low-Rank Adaptation (QLoRA) to ensure optimal performance on consumer-grade hardware. A unique aspect of our platform is the encoding of generated SLIs and SLOs as NFT objects, which are then stored on a Blockchain. This feature provides immutable record-keeping and facilitates easy verification and auditing of the SRE metrics and objectives. It enhances the traceability and accountability of the SRE processes, offering a verifiable and transparent record of the system's performance standards. The proposed SRE-Llama platform prototype has been implemented with a use case featuring a customized Open5GS 5G Core.


\end{abstract}



\section{Introduction}

Modern communication networks depend heavily on robust and efficient software for optimal performance. Examples include 5G network slicing software \cite{vani-2024}, Software Defined Network (SDN) management software \cite{zhao-2023}, and complex communication network simulators \cite{Ran-2020}. These examples highlight the critical role of software performance in overall system efficacy. Consequently, there is a significant emphasis on developing platforms that facilitate the creation of reliable communication software. This approach aligns with the principles of Site Reliability Engineering (SRE), which assert that operational issues are fundamentally software problems and should be addressed with software-based solutions. Central to this methodology are Service Level Indicators (SLIs) and Service Level Objectives (SLOs). SLIs offer quantifiable metrics related to service quality, such as response time or availability, critical for assessing and enhancing the communication network performance~\cite{sre}. SLOs, on the other hand, are the desired targets for these measures, defining what is considered acceptable performance. The implementation and maintenance of these metrics are vital for the continuous monitoring and improvement of service reliability. However, SRE faces several challenges, particularly from a developer's perspective. One of the main challenges is the complexity in defining and managing SLIs and SLOs~\cite{sli-slo}. This process requires not only an in-depth understanding of the systems in question but also an ability to translate operational goals into quantifiable and actionable metrics. Additionally, the dynamic nature of cloud-native environments adds to the complexity, as developers must continuously adapt and update these metrics to align with the evolving systems. Another significant challenge is the gap between development and operational practices. Developers often focus primarily on feature development and may not have extensive expertise in operational metrics or the tools commonly used in SRE, such as Prometheus and Grafana~\cite{prometheus-grafana}. This discrepancy can result in operational inefficiencies and misalignment between system design and implementation. Moreover, managing and responding to the vast amount of data generated in cloud-native environments can be overwhelming. Therefore, properly configured monitoring tools, alerting systems, and interpretation of data to make informed decisions, require a level of expertise that developers might not always possess.

To address these challenges, we propose a novel platform, called ``SRE-Llama'', which integrates Blockchain, Federated Learning, Generative AI (notably, Meta's Llama-3 LLM)~\cite{llama-3, llm}, and Non-Fungible Tokens (NFTs). SRE-Llama is designed to streamline and automate the formulation and management of SLIs and SLOs, effectively bridging the gap between software development and operations. This integration aims to make SRE practices more accessible and practical for developers, facilitating a future where SRE is seamlessly integrated into software development and operational workflows. The proposed platform operates by capturing and storing the key metrics from cloud-native services (e.g., as containers running on Kubernetes environment) within a time-series database, such as Prometheus. These metrics are then leveraged by Federated Learning models to identify the most relevant and impactful SLI metrics for different services, while simultaneously addressing data privacy concerns and the decentralization of data sources.
Subsequently, the custom-trained Meta's Llama-3 LLM is utilized to precisely generate corresponding SLIs, SLOs, Error Budgets, and alert mechanisms based on these identified SLI metrics. This method guarantees not only the relevance and context-sensitivity of the SLOs but also significantly reduces the manual effort that is traditionally required in establishing and upholding these standards. Advancing further, the platform integrates an additional layer of intelligence by adopting the federated machine learning models. These models specialize in examining the current state of the SLOs in conjunction with real-time metrics data. A distinctive feature of our platform is its capability to encode the devised SLIs and SLOs as NFT objects~\cite{nft}, subsequently storing them on a Blockchain. This approach ensures tamper-proof record-keeping, simplifies the verification and audit processes, and augments the traceability and accountability within the SRE framework. It creates a verifiable and transparent documentation of the system’s performance benchmarks. The SRE-Llama prototype has been developed using custom-trained Meta's Llama-3 LLM's~\cite{llama-3} and Bassa-ML~\cite{bassa-ml} blockchain-enabled coordinator-less federated learning platform with Long Short-Term Memory (LSTM) Recurrent Neural Network (RNN). Functions such as the crafting of SLO definitions and alert generation are adeptly managed by Llama-3 LLM. Furthermore, SRE-Llama uses a novel NFT schema, s-528, that is specifically designed for representing SLI/SLO objects as NFT tokens to reinforce the integrity and accessibility of SRE data. The proposed SRE-Llama prototype has been implemented with a use case featuring a customized 5G Core, Open5GS. The main contributions are summarized as follows:

\begin{enumerate}[noitemsep,nolistsep]
    \item The proposed end-to-end automated SRE platform to enhance the efficiency and reliability of communication network software services.
    \item Utilize a custom-trained Llama-3 LLM for generating SLIs, SLOs and managing error budgets and alerts, based on the identified SLI metrics, thereby reducing manual effort and improving contextual accuracy.
    \item Propose a blockchain-enabled, coordinator-less federated learning mechanism for the identifying of SLI metrics and SLO violations.
    \item Introduce a novel model to represent SLI/SLO metrics as NFT tokens and design an extensible NFT schema (s-528) specifically for encapsulating SLI/SLO objects, ensuring immutability and ease of verification.
\end{enumerate}

The rest of the paper is structured as follows: Section 2 discusses the architecture of the platform. Section 3 elaborates on the platform's functionalities. Section 4 details performance evaluation, implementation, and testbed setup. Section 5 reviews related work. Finally, Section 6 summarizes the proposed platform and offers recommendations for future work.

\begin{figure}[t]
\centering{}
\includegraphics[width=3.0in]{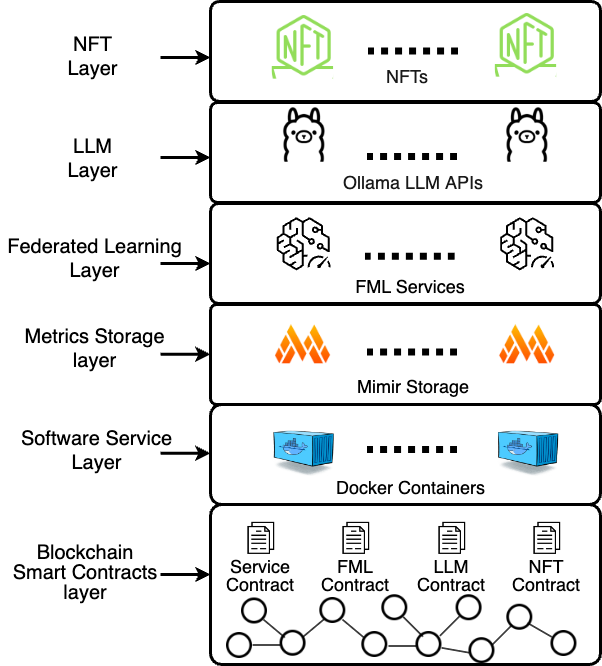}
\DeclareGraphicsExtensions.
\caption{Platform layered architecture.}
\vspace{-0.2in}
\label{indy528-architecture}
\end{figure}

\section{System Architecture}

Figure~\ref{indy528-architecture} describes the architecture of the platform. The proposed platform is composed of six layers: 1) Blockchain/Smart Contract Layer, 2) Software Service layer, 3) Metrics Storage Layer, 4) Federated Learning Layer, 5) LLM layer, and 6) NFT Layer. Each layer is briefly discussed in the following subesctions. 

\subsection{Blockchain and Smart Contracts Layer}
The Blockchain or Smart Contracts Layer orchestrates and governs all critical functions and coordination mechanisms within the system. At the core of this layer is the storage of all data, which includes a wide array of information such as user identities, container details, federated machine learning models, NFT objects, and data provenance information. This comprehensive data repository enhances the platform's reliability, security, and efficiency, ensuring all interactions and transactions are transparent and immutable. The architecture of this layer is built around five main smart contracts~\cite{rahasak}, each designed to manage specific aspects of the system: 1) \textit{Identity Contract}: This contract is responsible for managing user identities. It ensures secure and verifiable identification of users, which is fundamental for access control and interaction within the platform. 2) \textit{Service Registry Contract}: This contract is responsible for storing all the information(e.g., container information, deployments, metrics endpoints, etc.) regarding cloud-native software services (e.g., microservices) deployed within the platform, such as in a service registry. These services can be deployed in environments like Docker or Kubernetes~\cite{cloud-native-devops}. 3) \textit{Federated Learning Contract}: At the core of the platform's learning capabilities, this contract oversees all functions related to the federated learning process. It manages the distribution, execution, and aggregation of machine learning tasks across various nodes in a privacy-preserving manner. 4) \textit{LLM Contract}: Integral to the platform's operational objectives, this contract manages the SLIs, SLOs, including their generation and definition by utilizing the capabilities of custom-trained Llama-3 LLM~\cite{llama-3}. It ensures that SLIs and SLOs are dynamically updated and aligned with the evolving system parameters and performance metrics. 5) \textit{NFT Contract}: This contract handles the NFT-related functionalities. It manages the creation, storage, and transfer of NFT objects that represent the SLIs and SLOs, ensuring a secure and verifiable record of these critical metrics.

\subsection{Software Service Layer}
The Software Service Layer hosts all cloud-native software services. This layer is designed to support services deployed through Docker or Kubernetes-based container orchestration systems, which are widely recognized for their efficiency and scalability in managing containerized applications. In this layer, each software service, ranging from individual microservices to complex service compositions, is meticulously managed~\cite{container-deployment}. Crucial information about these services, including deployment details, metrics endpoints, and other operational data, is securely managed and stored in blockchain storage. This is achieved through the use of a dedicated Service Smart Contract, which ensures that all service-related data is handled with the utmost integrity, security, and accessibility. The metrics of these services will be collected and stored in time series data storages (e.g., Prometheus and Mimir).

\subsection{Metrics Storage Layer}
This layer serves as the central repository for all metrics collected from the various software services operating within the system. It is specifically designed to handle the vast and continuously evolving data generated by cloud-native services, capturing a comprehensive picture of their operational state. The metrics from the software services are systematically collected and stored in this layer~\cite{sre}. These metrics encompass a wide range of data points, including performance indicators, resource usage statistics, and operational health metrics, among others. The precise collection and storage of these metrics are critical for enabling real-time monitoring, performance analysis, and the effective management of service-level agreements. To manage this plethora of data, the Metrics Storage Layer utilizes time-series data storage systems, such as Prometheus and Mimir~\cite{prometheus-grafana, mimir}. These systems are particularly well-suited for handling time-stamped data and are capable of efficiently storing, querying, and managing large volumes of metrics over time. Their design is optimized for high-velocity data characteristic of modern, dynamic cloud-native environments, ensuring that the data is not only reliably stored but also readily accessible for analysis. The use of time-series databases like Prometheus and Mimir offer several advantages in this context. They provide powerful querying capabilities, enabling detailed analysis and insights into the performance and health of software services. Additionally, their scalability and robustness make them ideal for the system demands that must adapt to varying loads and rapidly changing conditions. The Federated Learning Layer builds ML models to detect SLIs of the services based on these metrics data. 

\subsection{Federated Learning Layer}
The Federated Learning Layer utilizes the metrics data stored in the Metrics Storage Layer to develop machine learning models in a privacy-preserving manner~\cite{bassa-ml}. This layer is specifically focused on leveraging federated learning techniques to discern and identify SLI metrics for the monitored services. In this layer, federated learning models are built using the metrics data accumulated from various software services. These models are intricately designed to analyze patterns and trends within the data, enabling them to effectively detect and pinpoint the key performance indicator (SLIs) metrics that are crucial for maintaining the operational integrity and efficiency of the services. A standout feature of this layer is the implementation of a blockchain-enabled, coordinator-less federated learning system~\cite{bassa-ml}. This approach addresses and mitigates potential vulnerabilities associated with centralized-coordinator-based federated learning systems, particularly those susceptible to attacks and single points of failure. By decentralizing the coordination mechanism and underpinning it with blockchain technology, the system enhances security, reliability, and trustworthiness in the model-building process. The orchestration of the layer is handled through the Federated Learning contract in the blockchain. 

\subsection{LLM Layer}
\begin{figure}[t]
\centering{}
\includegraphics[width=3.2in]{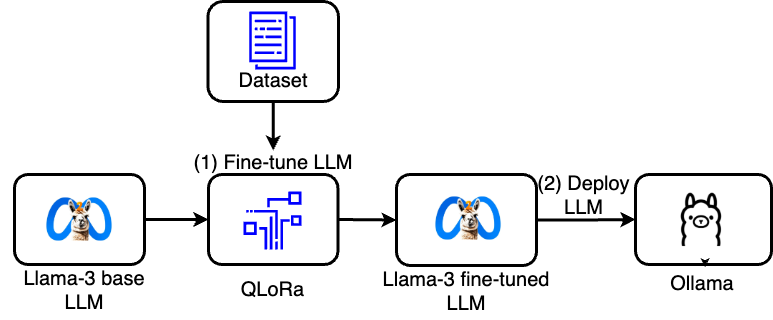}
\DeclareGraphicsExtensions.
\vspace{-0.1in}
\caption{Fine-tune LLM with QLoRA and deploy with Ollama.}
\label{llm-fine-tune}
\end{figure}

The LLM Layer is an integral component of the proposed SRE platform, refining the SLIs, SLOs, and error budgets for cloud-native services. This layer leverages the custom-trained Meta's Llama-3 LLM~\cite{llama-3}, to process and interpret the SLI metrics identified by the federated learning models. Functioning as the analytical and generative brain of the platform, the LLM Layer focuses on analyzing the SLI metrics generated through the federated learning models. Its primary objective is to define precise and contextually relevant SLIs, SLOs, error budgets, alerts (e.g., as PromQL queries~\cite{prometheus-grafana}), etc., for the various services being monitored. This process involves understanding the intricacies of each SLI, including its implications for service performance and reliability. The Llama-3 LLM is specifically fine-tuned/trained to excel in the task of PromQL query generation for SLOs, error budgets, alerts, etc. To fine-tune the Llama-3 LLM, we've undertaken a meticulous training process, collaborating with QLoRA with 4-bit quantized pre-trained language model into Low Rank Adapters (LoRA)~\cite{qlora}, as shown in Fig.~\ref{llm-fine-tune}. During the training phase, we meticulously feed the model with a rich dataset, encompassing SLIs, SLOs, error budgets, alerts, etc and the desired PromQL queries. The fine-tuned Llama-3 LLM run on Ollam~\cite{ollama}. The interaction with Meta's Llama-3 LLM is facilitated by the LLM smart contract through Ollam's LLM API. One of the key functionalities of this layer is instructing the LLM to generate SLOs, error budgets and alerts, etc based on the SLI metrics. This is achieved via custom prompts, which guide the custom-trained Llama-3 LLM in understanding the specific requirements and context of each service, Figure ~\ref{prompt}. Upon receiving these prompts, the Llama-3 LLM employs its extensive language understanding and generation capabilities to produce SLIs, SLOs and error budgets that align with the provided SLI metrics.



\subsection{NFT layer}
In this layer, all the SLIs and SLOs are encoded as NFTs~\cite{nft} and securely stored within the blockchain ledger. This approach not only innovates in the way service metrics are documented but also adds a layer of immutability and transparency to the process. The core functionality of this layer is to transform SLI and SLO information into NFT tokens. This encoding process ensures that each piece of information, being unique and immutable, is represented as a distinct digital asset on the blockchain. This method preserves the integrity of the data and makes it tamper-proof. To efficiently capture the nuances and details of SLIs and SLOs, we have developed a custom NFT schema, designated as s-528. This schema is specifically tailored to encapsulate all the relevant information about the SLIs and SLOs, ensuring that the NFTs are not only representative of the data but also structured in a way that facilitates easy access and interpretation. The blockchain component of the NFT Layer is governed by a specialized NFT smart contract. This contract manages the various functionalities associated with the NFT tokens, such as their creation, transfer, and querying. It acts as the backbone of the NFT Layer, providing a secure and efficient mechanism for handling the SLI and SLO NFTs.


\section{Platform Functionality}
There are six main functionalities of the platform 1) Service deployment, 2) Metrics capturing, 3) Federated Model building and SLI prediction, 4) SLO generation, 5) NFT tokenization, 6) Continuous monitoring and SLO violation prediction. This section goes into the specifics of these functions.

\subsection{Service Deployment}
The service deployment functionality represents the initial step in the proposed SRE-Llama platform, focusing on the deployment of cloud-native software services~\cite{container-deployment}. The source of the cloud-native software services will be built on CI/CD systems (e.g., GitHub-actions)~\cite{github-action}. Subsequently, the built services are containerized using technologies such as Docker, and are orchestrated and managed through container orchestration systems like Kubernetes. This approach not only ensures a high degree of scalability and flexibility but also aligns perfectly with modern practices in cloud-native application development. In this phase of service deployment, every aspect of the containerized services, including their configuration, deployment parameters, and runtime environments, is meticulously managed. The orchestration system, such as Kubernetes~\cite{container-deployment}, plays a crucial role in this process, handling the deployment, scaling, and operational aspects of these containerized services. This system ensures that the services are deployed efficiently, can be scaled according to the demands, and maintained with minimal downtime. 


\subsection{Metrics Collection}

Following the initial deployment of services, the next step in the SRE-Llama platform involves the collection of metrics from these services. This stage is vital for gaining insights into the performance and health of the deployed cloud-native applications. Metrics are scraped from various aspects of the services, capturing a wide range of data points such as performance metrics, resource utilization, response times, error rates, and more~\cite{sre}. The process of scraping these metrics is meticulously orchestrated to ensure comprehensive coverage and accuracy. The collected metrics provide a detailed snapshot of the operational state of each service, making them indispensable for effective monitoring and management. Once these metrics are scraped, they are stored in time-series databases, with Prometheus and Mimir being prominent choices due to their efficiency and scalability in handling time-stamped data. These time-series databases excel in storing, retrieving, and analyzing large volumes of metrics over time. 


\subsection{Federated Model Training and SLI Metrics prediction}

The subsequent phase in the platform's workflow involves predicting SLI metrics, driven by federated learning models trained on the data accumulated in the Metrics Storage Layer. In this process, federated learning models are trained utilizing the vast pool of metrics data. These models are adept at identifying patterns and trends within the data, which are indicative of key performance indicators crucial for the operational efficiency of the services. What sets our approach apart is the training of these models on a blockchain-enabled, coordinator-less federated learning system~\cite{bassa-ml}. This state-of-the-art system not only bolsters the security and privacy of the data but also mitigates the risks associated with centralized coordinator-based systems. By decentralizing the learning process, we enhance the robustness and reliability of the model training phase. These models play a crucial role in predicting the SLI metrics for various services~\cite{sli-slo}. By analyzing the metrics data, they can accurately forecast performance trends and operational health indicators, providing essential insights for SRE teams.

\subsection{SLO, Error Budget, Alert Generation}

Building upon the SLI metrics predicted by federated learning models, the next step involves generating SLIs, SLOs, error budgets, and alerts. This task is adeptly managed by Generative AI, specifically through the use of Meta's custom-trained Llama-3 LLM~\cite{llama-3}. At this juncture, the platform harnesses the fine-tuned Llama-3 to intelligently generate precise SLIs and SLOs, calculate error budgets, and configure alert mechanisms tailored to the specific needs and performance metrics of each service. The process of generating these SLOs and related elements begins with the crafting of custom prompts, as shown in Fig.~\ref{prompt}. These prompts are meticulously designed to guide Llama-3 in interpreting the SLI metrics, which have been identified through the federated learning models. The prompts encapsulate the nuances of each SLI metric and the operational context of the services, thus enabling Llama-3 to generate SLIs and SLOs that are not only data-driven but also contextually relevant and operationally feasible. Once the LLM receives the custom prompts, it applies its extensive language understanding and generation capabilities to produce SLIs, SLOs, error budgets, and alerts(e.g as PromQl queries) that align with the provided SLI metrics~\cite{sre}. 




\subsection{NFT Token Generation}


In this phase, the meticulously identified SLIs and intelligently generated SLOs are encoded as unique NFTs using the s-528 schema~\cite{indy528}. This schema is specifically designed to encapsulate the detailed information of SLIs and SLOs, ensuring that each token represents a comprehensive and immutable record of these crucial metrics. The tokenization process not only secures the integrity of these service metrics but also enhances their traceability and verifiability. Once tokenized, these NFTs are stored on the blockchain, providing a decentralized and tamper-proof ledger of the service performance standards. This storage method is pivotal as it ensures that the SLIs and SLOs are preserved in a form that is not only secure against unauthorized modifications but is also easily accessible for verification and auditing purposes~\cite{indy528}.



\subsection{Continuous Monitoring and SLO Violation Prediction}

The final step in the SRE-Llama platform focuses on continuous monitoring, a critical component that ensures the ongoing health and reliability of cloud-native services. This phase leverages the generated SLIs and SLOs to maintain vigilant surveillance over the services. The platform is designed to proactively monitor these metrics, ensuring that any deviations from the established standards are quickly identified and addressed~\cite{cloud-native-devops}. A key feature of this continuous monitoring process is the alerting system. The platform is configured to issue alerts when services are at risk of breaching their SLOs or exhausting their error budgets. These alerts are not just reactive notifications; they are anticipatory signals designed to provide sufficient lead time for teams to take corrective actions, thereby preventing potential service degradation or outages. Going a step further, the platform integrates an additional layer of intelligence through another set of federated machine learning models. These models specialize in examining the current state of the SLOs in conjunction with real-time metrics data. Their purpose is to predict whether the SLOs are likely to be violated in the near future.

\section{Implementation, Testbed Setup and Evaluation}


\begin{figure}[t]
\centering{}
\includegraphics[width=3.5in]{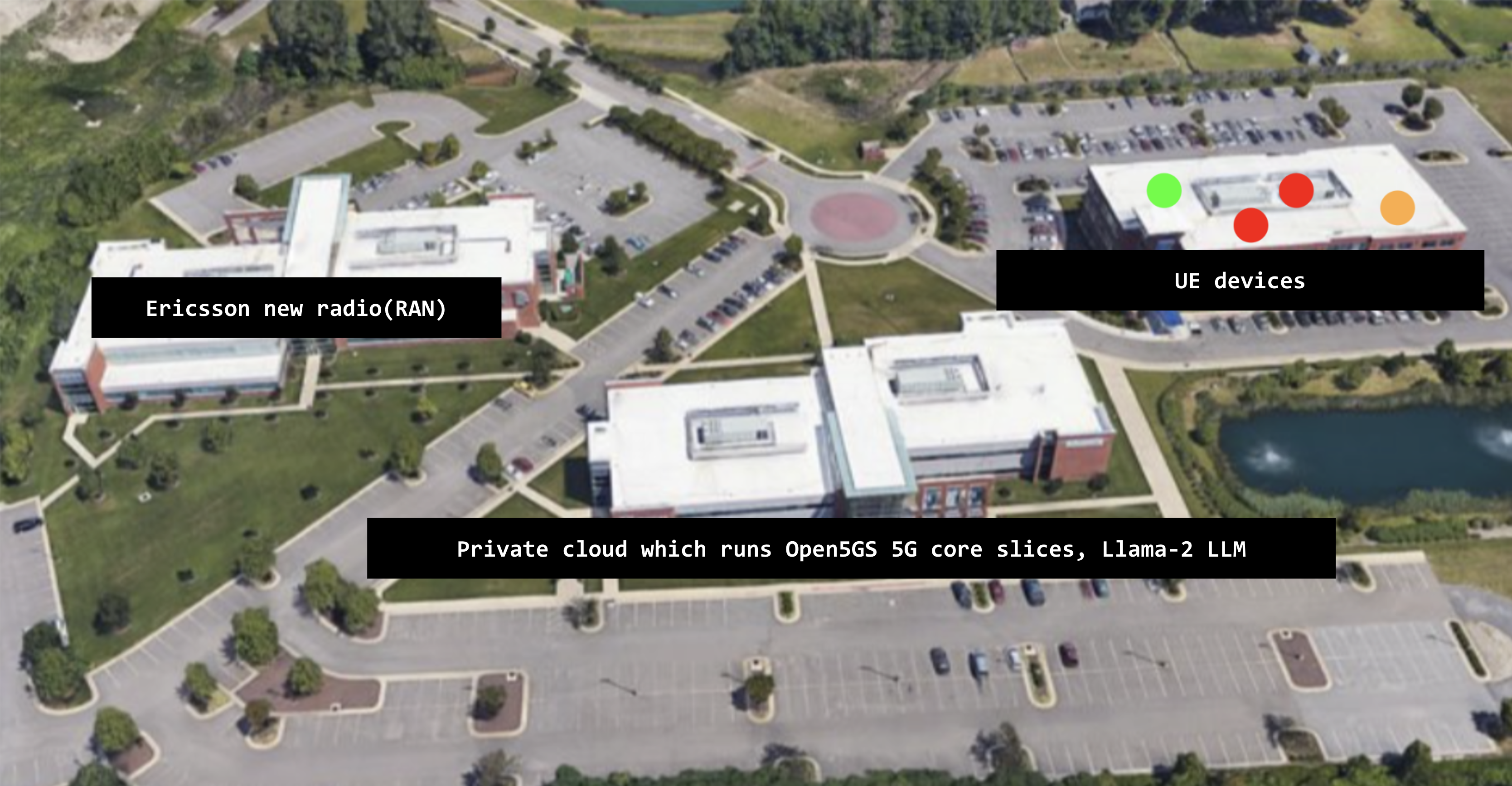}
\vspace{-0.2in}
\DeclareGraphicsExtensions.
\caption{Proposed large-scale testbed architecture with Ericsson's new RAN, Open5GS 5G-core, and on-prem Llama-3 LLM in VMASC Virginia US.}
\vspace{-0.2in}
\label{tesbed}
\end{figure}

The proposed SRE-Llama platform prototype and testbed has been implemented with a use case featuring a customized Open5GS 5G Core with Ericsson's new Radio in VMASC Virginia US(Fig.~\ref{tesbed}). The microservices of Open5GS are deployed using a Docker/Kubernetes-based container orchestration system~\cite{container-deployment}. Metrics from the customized Open5GS core microservices are captured and stored in Prometheus and Mimir~\cite{mimir}. Alert management is adeptly handled by Prometheus Alertmanager, while Grafana is utilized for dashboard setups. GitHub Actions~\cite{github-action} have been employed for continuous integration and deployment (CI/CD). The federated learning functions are implemented using the Bassa-ML blockchain-enabled, coordinator-less federated learning system~\cite{rahasak, bassa-ml} with LSTM~\cite{lstm} using TensorFlow~\cite{tensorflow} library. Additionally, the LLM layer, powered by the Llama-3-8B model, is used for generating SLOs and providing predictive insights based on analyzed metrics, enhancing the platform’s decision-making and automation processes. The Llama-3-8B quantized LLM runs on a consumer-grade server without using a GPU. The fine-tuning/training process of the LLM is conducted through QLoRA with a 4-bit quantized pre-trained language model adapted into Low Rank Adapters (LoRA)~\cite{qlora}. NFT tokens are structured through JSON encoding, adhering to the s-528 token schema. The platform’s performance is evaluated in two key areas: LLM and blockchain-enabled federated learning.

\begin{figure}[t]
\centering{}
\includegraphics[width=3.4in]{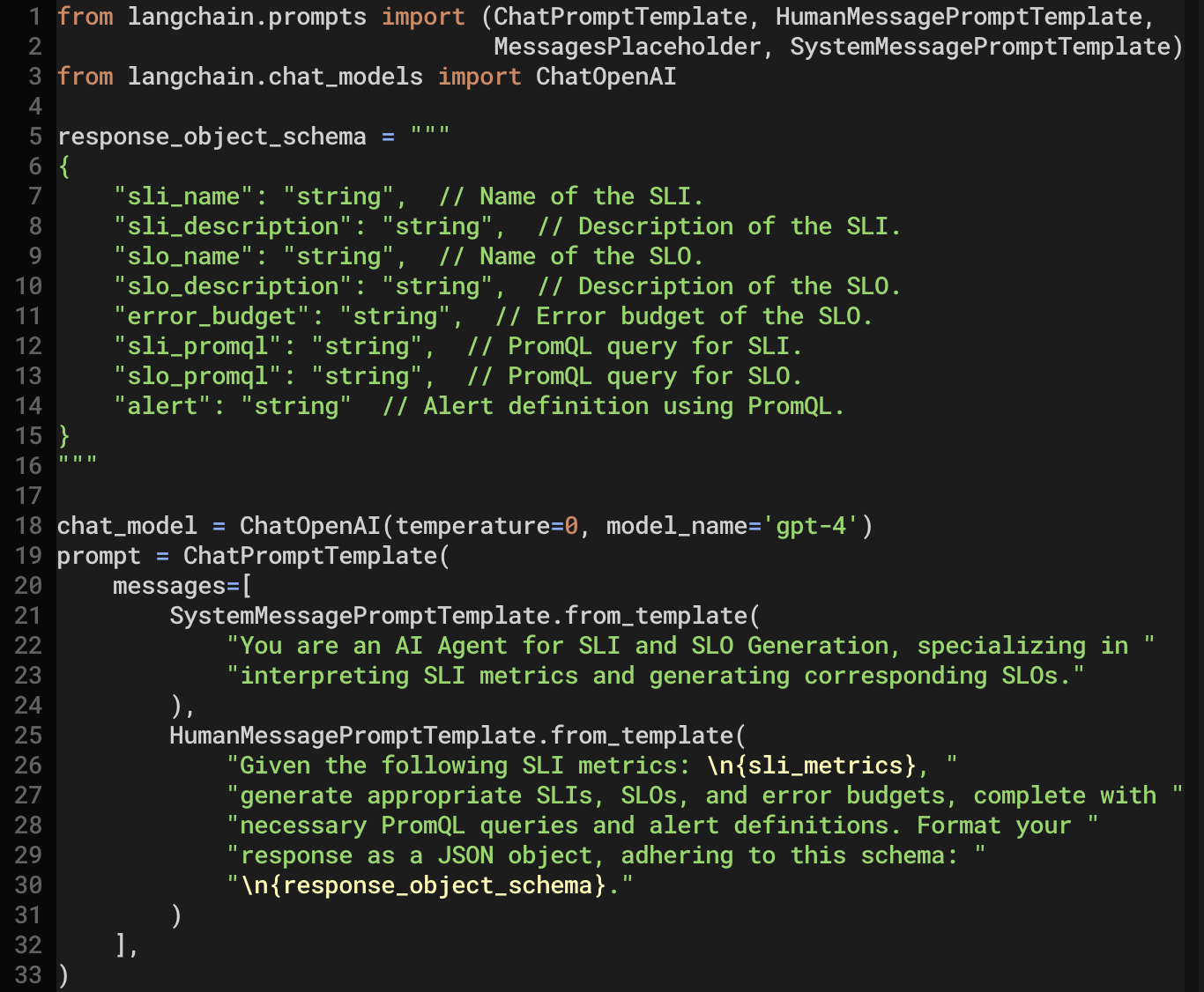}
\DeclareGraphicsExtensions.
\caption{SLO generation prompt.}
\label{prompt}
\end{figure}

\begin{figure}[t]
\centering{}
\vspace{-0.1in}
\includegraphics[width=3.4in]{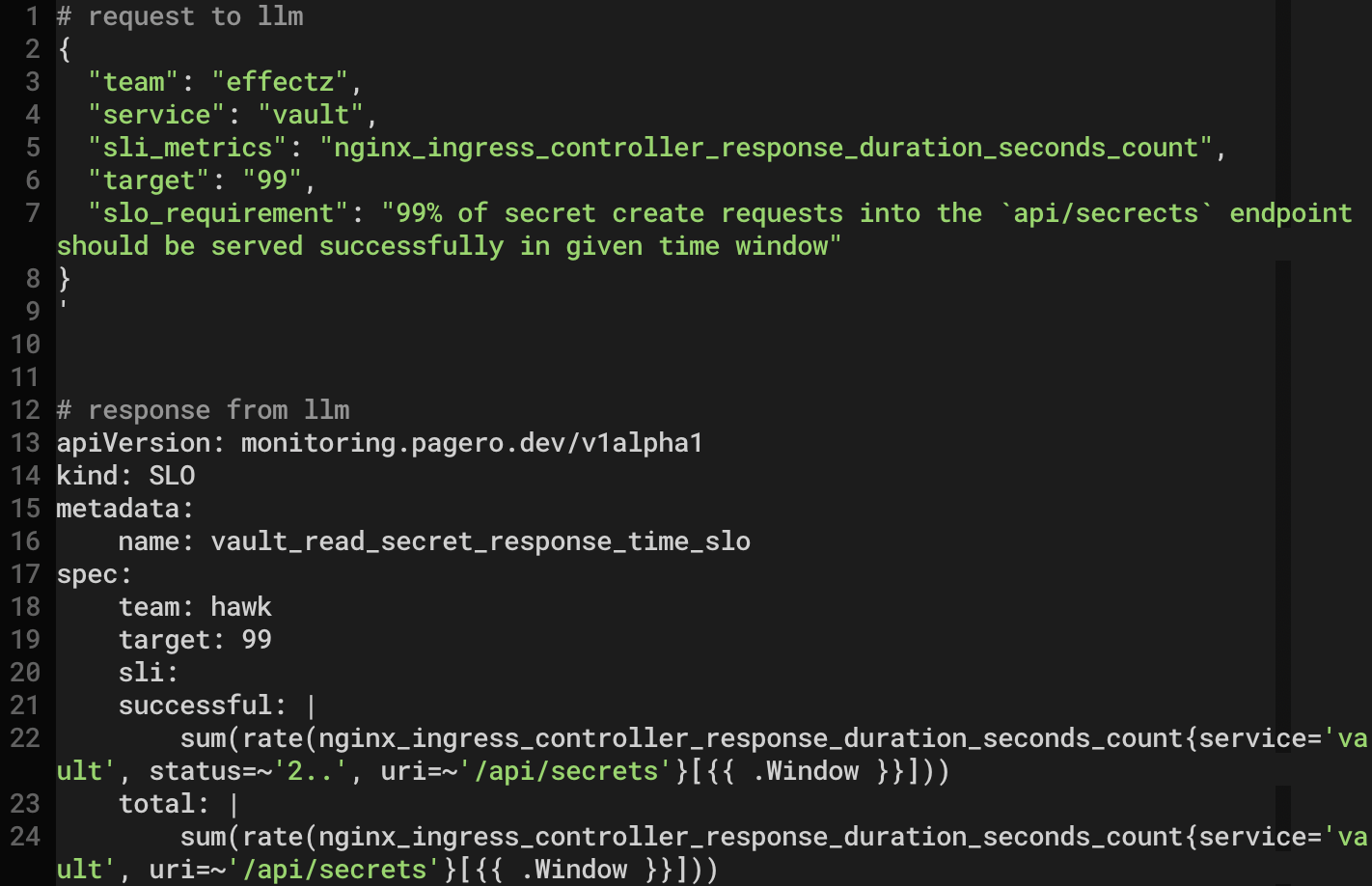}
\DeclareGraphicsExtensions.
\caption{SLO for the secret creation request.}
\label{slo1}
\end{figure}

\begin{figure}[t]
\centering{}
\vspace{-0.1in}
\includegraphics[width=3.4in]{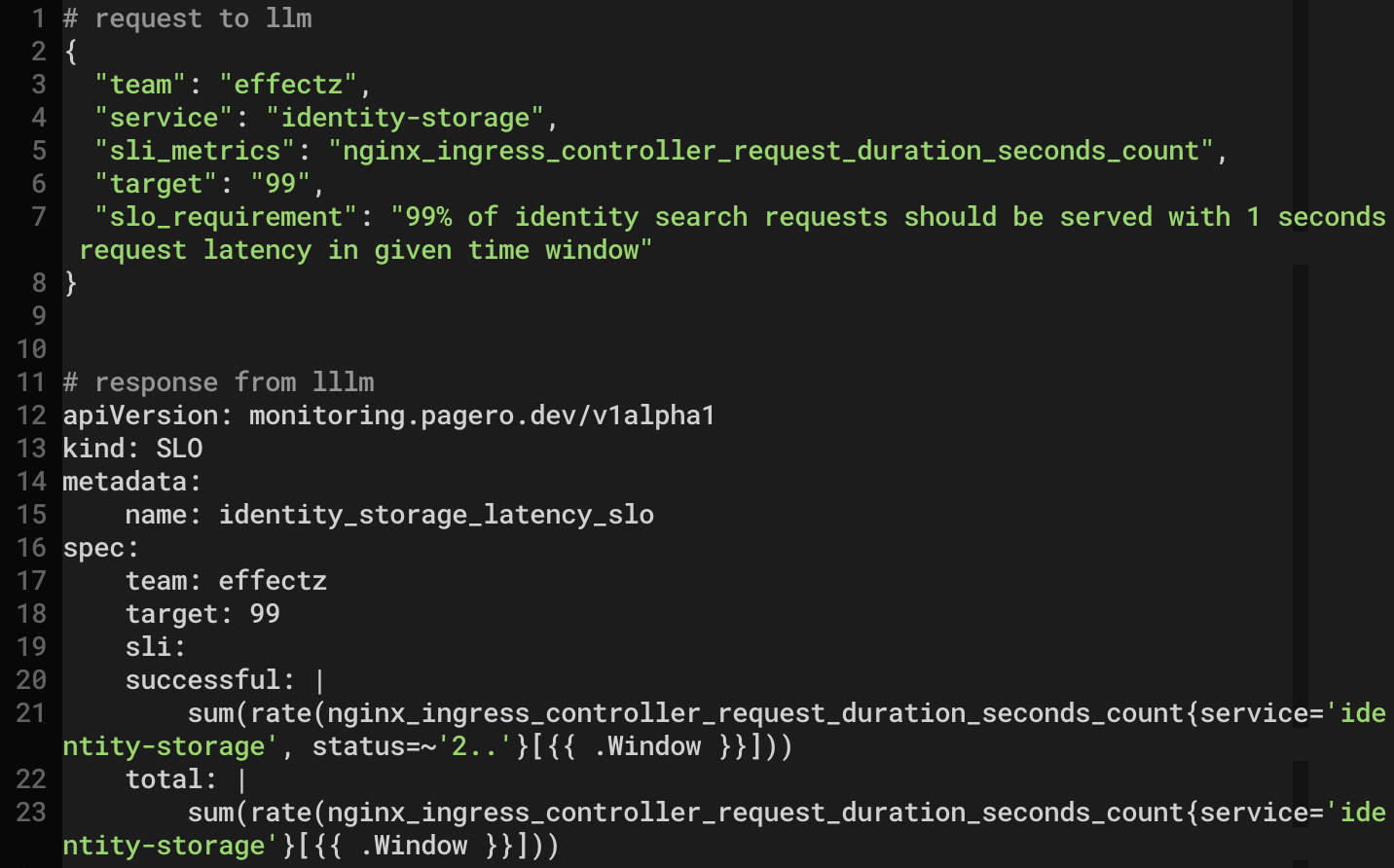}
\DeclareGraphicsExtensions.
\caption{SLO for the identity search request.}
\label{slo2}
\end{figure}

In the LLM evaluation, we analyzed the responses from the LLM regarding its ability to generate SLO objects. Prompts guide the custom-trained Llama-3 LLM in understanding the specific requirements and context of each service. Figure ~\ref{prompt} illustrates an example prompt used on the SRE-Llama platform to instruct the LLM in generating SLOs. Figure~\ref{slo1} presents an example SLO generation scenario for the secret creation request of the ``Vault" service in the customized Open5GS 5G core. It instructs the generation of an SLO where 99\% of secret creation requests to the `api/secrets` endpoint should be successfully served within a given time window, using the respective SLI metrics. Figure~\ref{slo2} presents another example of SLO generation for identity search requests at a service called ``Identity-Storage". It instructs the generation of an SLO where 99\% of document search requests should be served within 1 second of request latency in the given time window, according to the respective SLI metrics.

\begin{figure}[t]
\centering{}
\vspace{-0.2in}
\includegraphics[width=2.8in]{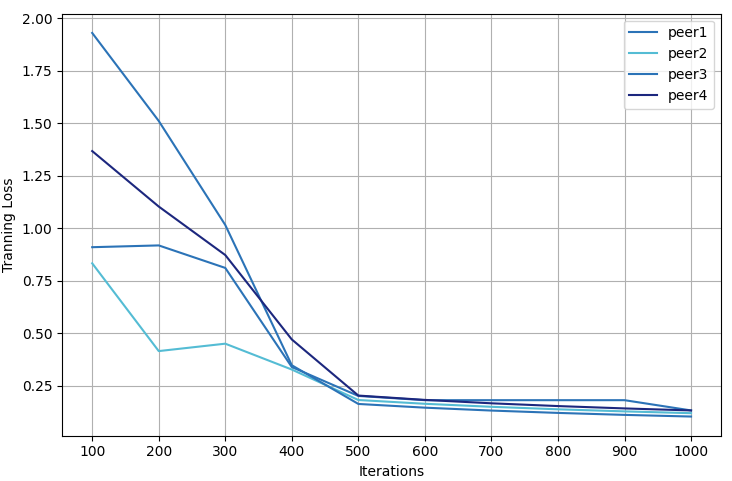}
\DeclareGraphicsExtensions.
\caption{Federated model training loss in different peers}
\vspace{-0.1in}
\label{federated-tranning-loss}
\vspace{-0.2in}
\end{figure}

\begin{figure}[t]
\centering{}
\vspace{-0.1in}
\includegraphics[width=2.8in]{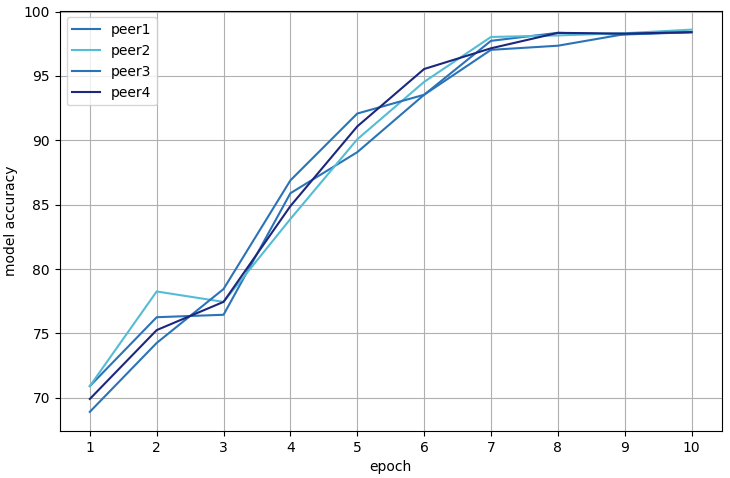}
\DeclareGraphicsExtensions.
\vspace{-0.1in}
\caption{Federated model accuracy in different peers}
\label{federated-tranning-accuracy}
\end{figure}

\begin{figure}[t]
\centering{}
\includegraphics[width=2.8in]{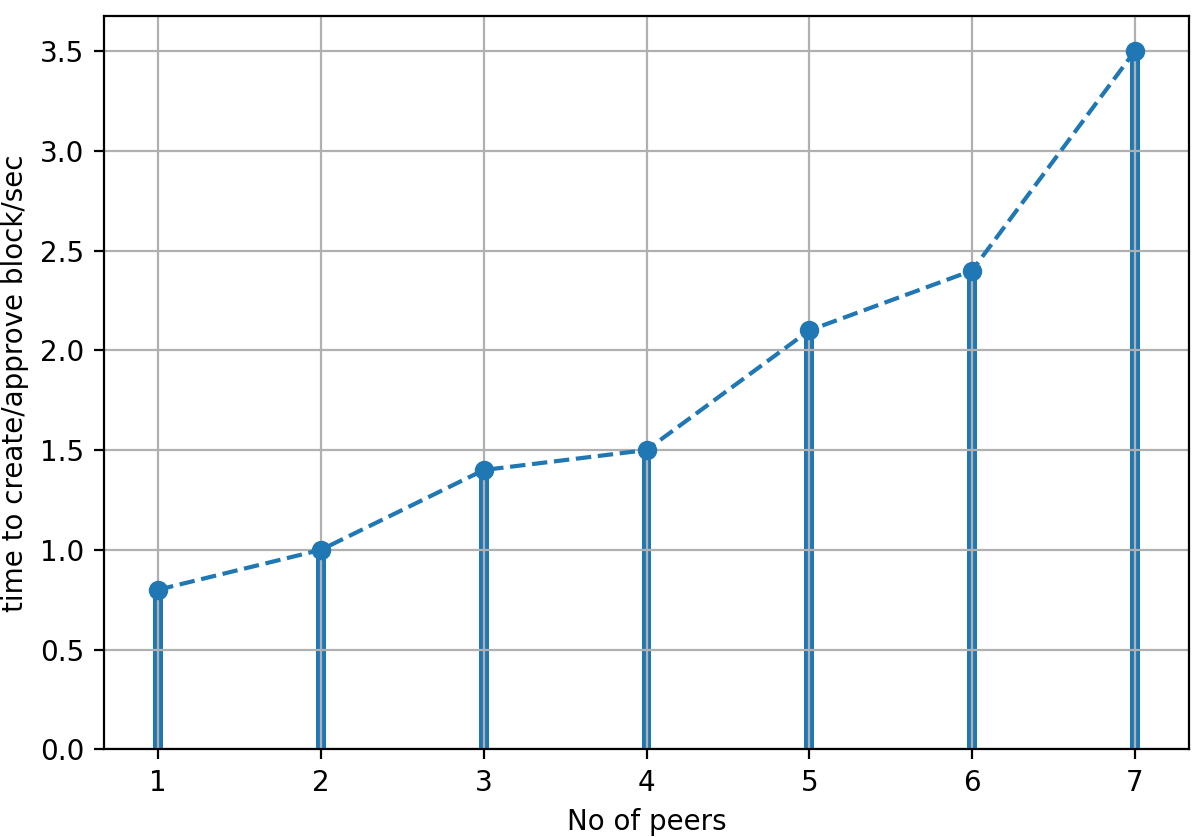}
\DeclareGraphicsExtensions.
\caption{Block creation time and \# of blockchain peers}
\vspace{-0.2in}
\label{block-time}
\end{figure}

In the evaluation of blockchain-enabled federated learning, we focused on assessing the accuracy and training loss of federated learning models, as well as the performance of the blockchain system. The federated learning process involved numerous iterations to refine the model's accuracy. In this evaluation, we trained the model over 1,000 iterations and plotted both the accuracy and training loss. Figure~\ref{federated-tranning-loss} illustrates the variation in total training loss across different peers in each iteration. Figure~\ref{federated-tranning-accuracy} displays the accuracy of the federated machine learning model. Additionally, the block generation time, measured while increasing the number of peers up to seven, is shown in Figure~\ref{block-time}. 

\begin{table*}[!htb]\centering
\caption {Platform comparison}
\vspace{0.1in}
\begin{adjustbox}{width=1\textwidth}
\label{t_bc_platforms}
\begin{tabular}{cccccccccc}
\toprule

\thead{Platform} & \thead{Centralized/\\Distributed} & \thead{Blockchain\\Support} & \thead{Running\\Blockchain} & \thead{Supported RMF\\Frameworks} & \thead{Data Provenance\\Support} & \thead{NFT\\Support} & \thead{Continuous ATO\\ Support} & \thead{AI\\Integration} \\
\midrule

SRE-Llama & Distributed & \cmark & \makecell{Bassa-ML} & \makecell{PCI DSS, NIST 800, STIG} & \cmark & \cmark & \cmark & \cmark\\

Contineous RMF~\cite{contineous-rmf} & Distributed & \cmark & \makecell{N/A} & \makecell{NIST} & \xmark & \xmark & \xmark & \xmark\\

\makecell{Perspective RMF}~\cite{perspective-rmf} & Distributed & \cmark & \makecell{N/A} & \makecell{N/A} & \xmark & \xmark & \xmark & \xmark\\

Letstrace~\cite{letstrace} & Distributed & \cmark & \makecell{Rahasak} & \maketitle{N/A} & \cmark & \xmark & \xmark & \xmark\\

Vind~\cite{vind} & Distributed & \cmark & \makecell{Rahasak} & \makecell{N/A} & \cmark & \xmark & \xmark & \xmark\\

SmartGrid-RMF~\cite{nist-smart-grid-rmf} & Distributed & \xmark & \makecell{N/A} & ~\makecell{NIST} & \cmark & \xmark & \cmark & \xmark\\
\bottomrule
\end{tabular}
\end{adjustbox}
\end{table*}

\section{Related Work}

Numerous researchers have focused on enhancing the security and reliability of cloud-native software applications by integrating AI, blockchain, and other technologies. This section outlines and summarizes the key elements and features of these research efforts, as detailed in Table ~\ref{t_bc_platforms}. 

``Continuous Cybersecurity Management Through Blockchain Technology"~\cite{contineous-rmf} introduces a novel approach utilizing blockchain technology to address delays in product release cycles and enhance security and functionality. The paper "Perspectives on risks and standards that affect the requirements engineering of blockchain technology"~\cite{perspective-rmf} explores the potential of blockchain technology to revolutionize business transactions by introducing a trust model based on algorithms. ``Letstrace"~\cite{letstrace} presents a blockchain-based cyber supply chain provenance platform that integrates TUF and In-ToTo frameworks, verifying software updates and enhancing supply chain security. It proposes blockchain-enabled federated learning to analyze cyber supply chain information, fortifying the overall security posture and enabling further analytics. In ``Vind"~\cite{vind}, a blockchain-based cyber supply chain provenance platform is proposed to address vulnerabilities in the Energy Delivery Systems supply chain. This platform quantifies risk metrics, assesses attack severity, and influences mitigation strategies, enhancing Industrial Control Systems hardware and software security. The paper "Blockchain Technology for Smart Grids: Decentralized NIST Conceptual Model"~\cite{nist-smart-grid-rmf} evaluates the impact of blockchain on decentralizing smart grids using the NIST conceptual model. It assesses blockchain's applicability, guiding smart grid developers and researchers in implementing decentralized blockchain-enabled smart grids.

\section{Conclusions and Future Work}

To address the need for a platform to produce reliable software for communications and networking systems, we proposed the novel SRE-Llama platform. By integrating Blockchain, Federated Learning, Generative AI (specifically Meta's Llama-3 LLM), container orchestration systems like Kubernetes, and time-series databases like Prometheus and Mimir, this platform represents a novel approach to managing cloud-native services. Our platform addresses key challenges in SRE, including the automation of service deployment, efficient metrics capturing, intelligent SLI metrics prediction and dynamic SLI, SLO generation. The integration of a blockchain-enabled, coordinator-less federated learning system for model training enhances the security and decentralization of data processing. The utilization of Meta's Llama-3 LLM for generating SLOs and error budgets based on SLIs illustrates the platform's capability in leveraging Generative AI to enhance service management and reliability. The NFT tokenization of SLIs and SLOs on the blockchain is a pioneering feature, ensuring the immutability, traceability, and verification of critical service metrics. This step not only enhances the auditability of SRE processes but also sets a new standard in the transparency and accountability of cloud-native service management. Lastly, the continuous monitoring mechanism, fortified with predictive capabilities through federated learning models, shifts the focus from reactive to proactive management of services. This ensures high reliability and customer satisfaction, aligning with the core objectives of SRE. In conclusion, the proposed SRE-Llama platform offers a scalable and forward-looking solution for effectively managing the increasingly complex landscape of communications and networking software systems. The implementation of the SRE-Llama testbed, featuring a customized Open5GS 5G core with Ericsson RAN, serves as a strong endorsement of the platform's capability to enhance operational reliability and efficiency in real-world environments. In our future work, we intend to integrate a variety of open-source LLMs to further enhance the capabilities of the SRE-Llama platform.




\bibliographystyle{IEEEtran}
\bibliography{references}

\end{document}